\documentclass{article}

\usepackage{cite}
\usepackage{subfigure}
\usepackage{geometry}
\usepackage{hyperref}
\usepackage{graphicx}
\usepackage{indentfirst}
\usepackage{amsmath}
\usepackage{bm}

\title{Effect of baffles on pressurization and thermal stratification in a LN2 tank under micro-gravity}
\author{Zuo Zhongqi, Jiang Wenbing, Huang Yonghua${}^*$}
\date{}

\begin{document}
\maketitle

\section*{ABSTRACT}
Researches on the impact of existing baffles on sloshing suppression of two-phase fluids in storage tanks have been widely 
conducted in literature. However, few studies focus on the effect of the baffles on self-pressurization or thermal stratification 
of the fluids in containers. This paper uses Volume of Fluid (VOF) method to simulate the thermodynamic and fluid dynamic behavior
 of liquid nitrogen in a tank with different baffle structures under microgravity environment. Groups of gravity levels, fill levels 
 and distances, angles and gaps of baffles, are compared and analyzed. Up to 54\% difference in pressurization can be observed by optimizing 
 the baffle structure and metrics, which is significant to achieve the highest performance of storage fluid control in the tank.  

\vspace{0.5cm}

{\bf Keyword:} Baffles, Self-pressurization, Thermal stratification, Microgravity

\section{INTRODUCTION}

Liquid hydrogen or oxygen fuel tanks are important parts in a space vehicle. Baffle structures have been widely adopted to avoid liquid
 sloshing in these tanks no matter on ground in normal gravity or on orbit in microgravity. The baffles are very effective in mitigating 
 strong shaking of liquid when the tank is impacted by a sudden external force. There are many experiments and simulations performed in normal gravity. 

Panzarella\cite{panzarella2003validity} states that researches on cryogenic fluid storage in tanks generally concern three aspects: self-pressurization,
 fluid flow/convection and thermal stratification, and behavior and evolution of liquid-vapor interface. Sloshing behavior in a tank has been 
 studied since 1960s\cite{sumner1966experimental}. Chintalapati\cite{chintalapati2008parametric} found that baffles mitigate the peak slosh height
  and if there is a small hole on baffle, the peak slosh height will increase 10-25\%. Hasheminejad \cite{hasheminejad2011effect}\cite{hasheminejad2012sloshing}investigated
   the transverse two-dimensional sloshing modes in both circular and elliptical tanks using a simple semi-analytic approach based on linearized theory.
    The tank is partially filled and have a vertical baffle stand in the bottom of the tank. Yoon\cite{yoon2015effect} investigated effect of baffles on
    mitigating sloshing with arrays of holes on the baffle. Behavior of propellant in a baffled tank in microgravity is attacking more researchers' concern, 
    including NASA\cite{dodge2000new}. Kannapel\cite{kannapel1987liquid} analyzed sloshing behavior of liquid oxygen in space shuttle external tanks.
     All these researches shows the important roles of baffles in tank design to reduce sloshing as well as mitigate impact force to the supporting structure of the tank.

Other than the conditions with acceleration, baffles could also be useful to stationary storage of fluids in containers in perspective of hydrodynamics
and thermodynamics. However, researches in this area is far more imperfect compared to dynamic study of baffles. Panzarella\cite{panzarella2003validity} 
systematically presented an analysis of self-pressurization process of cryogenics in a tank in normal gravity. Behruzi\cite{behruzi2014cryogenic} studied
 the temperature difference during sloshing via both experiment and simulation, and bring up a simplified model to describe heat transfer in sloshing. 
 But his model did not include a baffle structure. Ma\cite{ma2017investigation} used CFD to investigate the no-vent filling process in a liquid hydrogen 
 tank. Her study shows that the mixing in microgravity is more adequate, and results in a more steady pressure behavior. Adam\cite{adam2014design} proposed a baffled tank used for aerial vehicles, whose baffles have multilayer structure. He considered the heat leak to the tank, but failed to concern the effect of heat leak to the fluid field in the tank. Grayson\cite{grayson2006cryogenic}
  conducted microgravity experiments on the effect of baffles on the convection of the ullage in the tank on the AS-203 satellite, and found that the 
  baffles can promote vortex in the ullage of the tank by changing the velocity vector. He also emphasized that the effect of natural convection cannot 
  be ignored in microgravity, although it is on low level. To our knowledge, there is a lack of investigation on how the baffles will affect the thermal
   stratification or pressurization of fluid in the tank. This paper focuses on the effect of different distances, angles and gaps of the baffles as well
    as the liquid fill levels and gravity levels on the temperature profiles and self-pressurization behavior of nitrogen in microgravity environment.
     Performances of baffles in different conditions are considered. A specialized structure can be constructed based on the results to have the highest 
     performance for pressure and velocity suppression, which can be helpful in designing space cryogenic containers. 

\section{COMPUTATION MODEL}

\subsection{Setup}
\begin{figure}[htb!]

\centering
\subfigure[Diagram of tank used in simulation]{
    \label{diagram}
    \includegraphics[height=6cm]{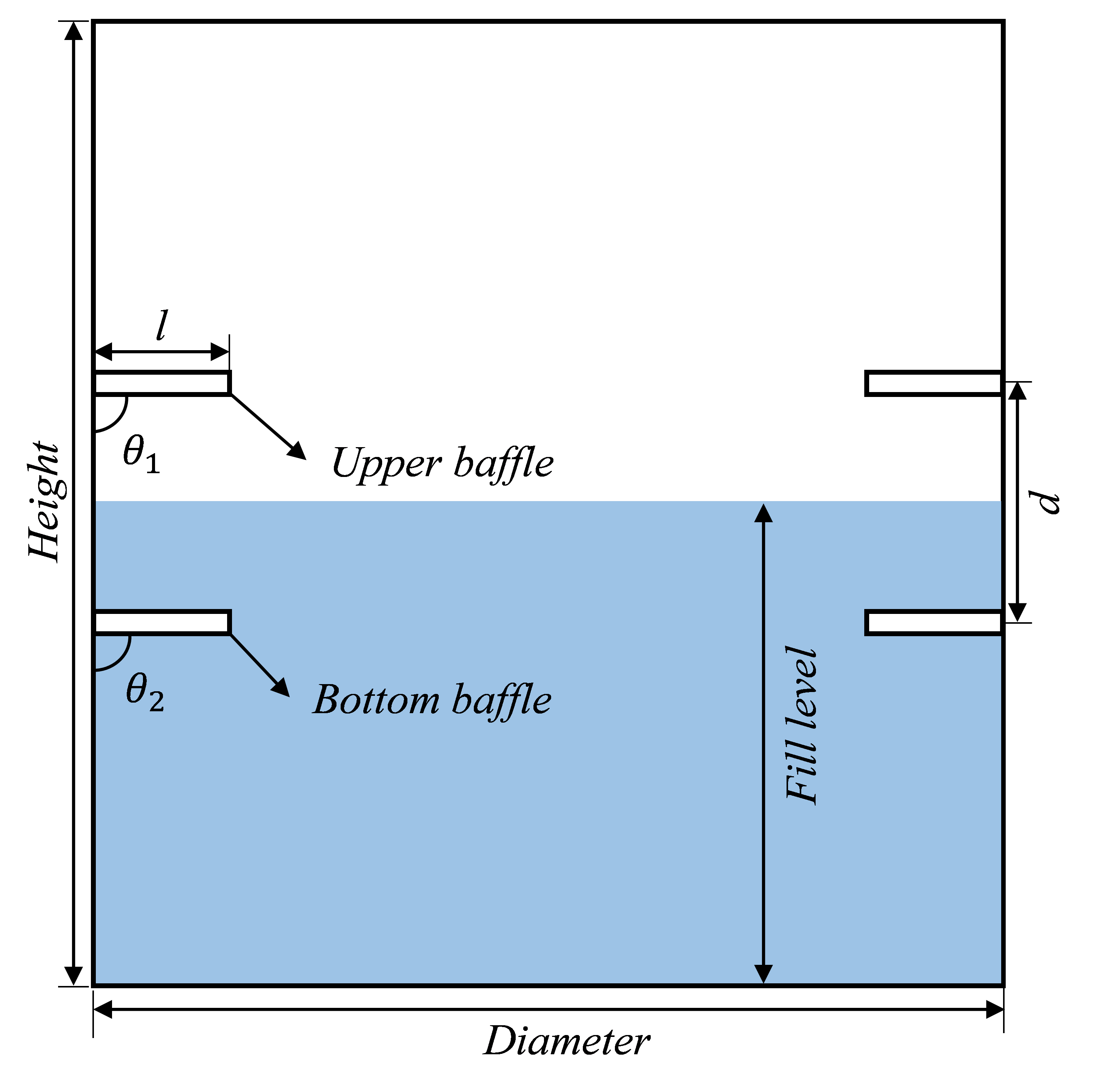}
}
\subfigure[Mesh example]{
    \label{mesh}
    \includegraphics[height=6cm]{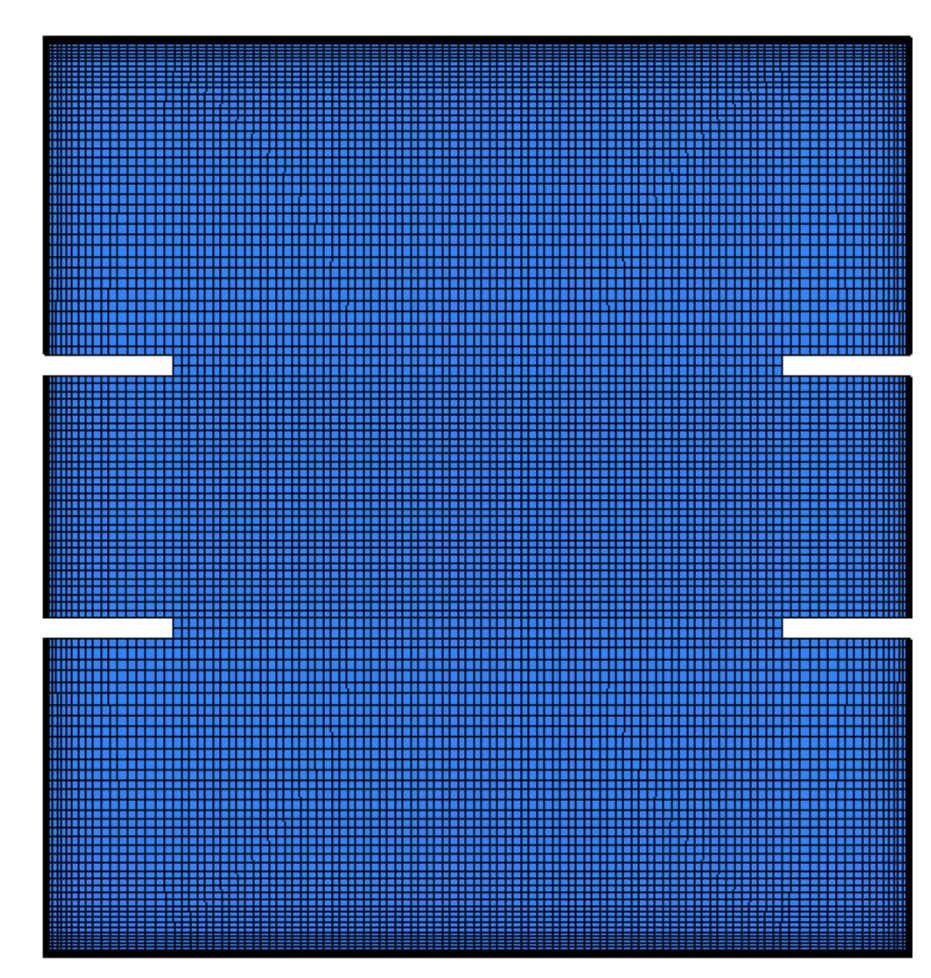}
}
\caption{Tank diagram and Mesh Example}
\label{s&m}

\end{figure}

The modeling tank is in cylindrical shape with diameter and height of 201 mm and 213 mm, respectively, which is in accordance with the one without baffles
 from reference \cite{seo2010analysis}. Simulations in 2D are conducted in various gravity environments and gravity level set includes $10^{-5} g_0$,$10^{-3}
  g_0$,$10^{-1} g_0$. It is reasonable because the tank structure is symmetrical along its axis. Six different fill levels are set to check the performance
   of baffles during the reduction of liquid. Annular plane baffles are installed on the internal side of the tank as shown in Fig.\ref{diagram}, where $d$
   is the distance between two baffles. The baffles are configured in the forms of single baffle and twin baffles with different included angles, where 
   $\theta$ is the angle between baffles and tank wall. The parameters $d$, $\theta$ and gaps are varied to optimize the highest performance for pressure 
   and velocity suppression by the baffles. All simulations in this paper are summarized in Table.\ref{pars}.

% test
\begin{table}
\centering
\begin{tabular}{cc}
\hline
 Parameters & Group of setups \\
 \hline
 Gravity level & $10^{-1}g_0$, $10^{-3}g_0$, $10^{-5}g_0$ \\
 Fill level & 30\%, 40\%, 50\%, 55\%, 60\%, 70\% \\
 Distance($d$) & 0mm, 20mm, 40mm, 60mm\\
 Orientation($\theta$) & 105\&75, 75\&105, 90\&90 \\
 Gap & 0mm, 5mm, 10mm\\
 \hline
 \end{tabular}
 \caption{Simulations executed in this paper}
 \label{pars}
 \end{table}

 VOF (Volume of Fluid) method is adopted to track the interface between liquid and gaseous nitrogen. Realizable k-$\epsilon$ model is used for the 
 turbulence. This model adds mathematical restrictions to the origin model and is widely used in relevant researches\cite{kartuzova2011modeling}. PISO scheme and Body Force 
 Weighted method for simulating pressure are selected to gain a larger time step. The evaporation and condensation models are applied, and the frequency 
 is set to be $ 5\times 10^{-3} $ and $1\times 10^{-5}$ respectively. The evaporation temperature under operating pressure is obtained from NIST.
  A uniform heat flow of $\mathrm{10 W/m^2}$ is applied to all sidewalls of the tank. The baffles are usually connected to the tank wall and increase the parasitic heat leakage.
  The total heat leakage is larger when the tank is mounted with baffles. Thus the baffles is treated with same heat flow of  $\mathrm{10 W/m^2}$.
 Liquid nitrogen is saturated at initialization.
VOF method\cite{hirt1981volume} introduces phase fraction $\alpha$ to each phase, and couples it with the governing equations. By solving the equations, 
fractions of each phase in each cell can be specified, and $\alpha = 0$ means pure gas, $\alpha=1$ means pure liquid in this case. Material derivative of
 $\alpha$ equals the mass transfer between liquid and gas, and sum of fraction in one cell equals 1, which is,
\begin{equation}
\frac{\partial \alpha_i}{\partial t} + u_j \frac{\partial \alpha_i}{\partial x_j} = \dot{m}
\end{equation}
\begin{equation}
 \sum \alpha_i = 1
\end{equation}
where $\dot{m}$ is the mass transfer.
Continuum surface force (CSF) model is used to model surface tension. This method adds a source term to the momentum equation, 
\begin{equation}
F_{vol} = \sum_{i<j} \sigma_{ij} \frac{\alpha_i \rho_i \kappa_j \bf{\nabla} \alpha_j + \alpha_j \rho_j \kappa_i \bf{\nabla} \alpha_i}{\frac{1}{2}(\rho_i +\rho_j)}
\end{equation}
where $\kappa$ is defined in terms of the divergence of surface normal, $\sigma$ is the surface tension.

Regarding the operation pressure and temperature, the fluids are treated as incompressible fluid, so the governing equations are,

\begin{equation}
\frac{\partial \rho}{\partial t} + \bf{\nabla} \cdot (\rho \bf{u}) = 0
\end{equation}

\begin{equation}
\rho \frac{\mathrm{D} \bf{u}}{\mathrm{D}t} = -\bf{\nabla} p + \mu \bf{\Delta} \bf{u} + \rho \bf{f}
\end{equation}

\begin{equation}
\rho \frac{\mathrm{D}}{\mathrm{D}t}(e+\frac{1}{2}\bf{u}\cdot\bf{u}) = \bf{\nabla} \cdot (\bf{\Sigma} \cdot \bf{u}) + \rho \bf{u} \cdot \bf{f} - \bf{\nabla} \cdot \bf{q} 
\end{equation}

Equation (4)-(6) are continuum equation, momentum equation and energy equation respectively, where $\bf{f}$ includes body forces like gravity 
force and source terms like CSF term, $\bf{q}$ is heat flux, $e$ is internal energy.

\subsection{Validation}

\begin{figure}[htb!]
\centering
\subfigure[Model validation]{
    \label{modelval}
    \includegraphics[width=7cm]{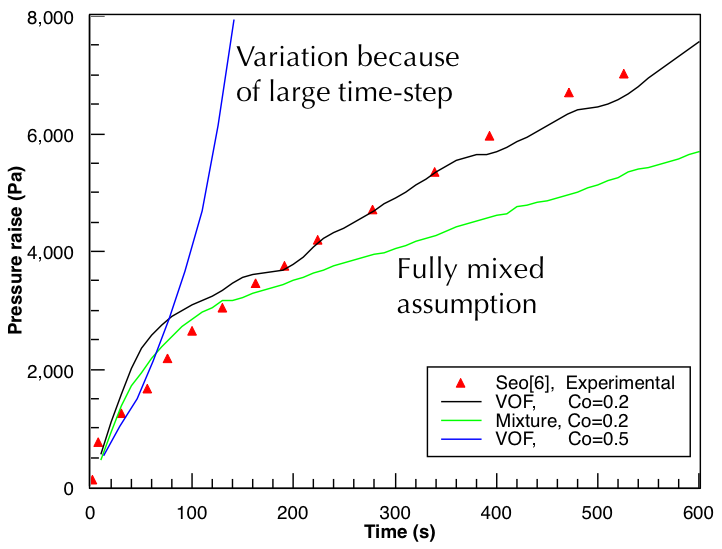}
}
\subfigure[Mesh independence check]{
    \label{meshindep}
    \includegraphics[width=7cm]{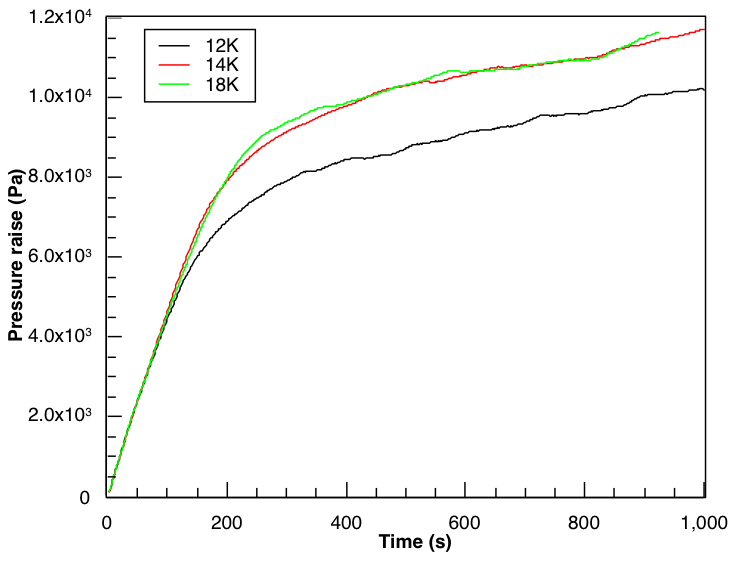}
}
\caption{Validation of simulation}
\label{mv&mi}
\end{figure}

Several cases are conducted to simulate the self-pressurization process of a LN2 tank under 1-g gravity. The results are shown in Fig.\ref{mv&mi}.
 The pressure will grow rapidly when Courant number equals 0.5 as shown in Fig.\ref{modelval}, representing a relatively large time step. However, 
 the prediction by VOF model with Courant number equaling 0.2 presents high accuracy against the experiment data by Ref\cite{seo2010analysis}.
  Mixture model predicts a lower pressure rise, which is related to the assumption in mixture model that phases are fully mixed. Fig.\ref{meshindep}
   is the check of mesh independence. A clear difference can be observed between 12K mesh and 14K mesh in predicting pressure under same condition,
    while difference between 14K mesh and 18K mesh can be ignored. Regarding the consumption of computation resources, 14K mesh is chosen to perform
    the simulations. Thus all simulations are operated with $Co=0.2$ and mesh number of around 14K. Example mesh in simulation is 
    shown in Fig.\ref{mesh}. Mesh is refined near the tank wall, and at least 2 layers of mesh is guaranteed at the baffles. The convergence criteria for 
for turbulent energy and dissipation rate is $10^{-4}$, for energy equation is $10^{-7}$, and for continuity equation is $10^{-3}$. 

\section{RESULTS AND DISCUSSIONS}
\subsection{Gravity level}
Theoretical analysis done by Dodge et al.\cite{dodge2000new} showed the pattern of interface shape in microgravity. The capillary area $A_c$, which is defined as,
\begin{equation}
A_c = A_i -cos\theta_c A_{wet}
\end{equation}

where $A_i$ is area of interface, $A_{wet}$ is wet wall area and $\theta_c$ is the contact angle between liquid and wall. Capillary area will reach its minimum
 at 0 gravity at equilibrium. 

\begin{figure}[htb!]
\centering
\includegraphics[width=10cm]{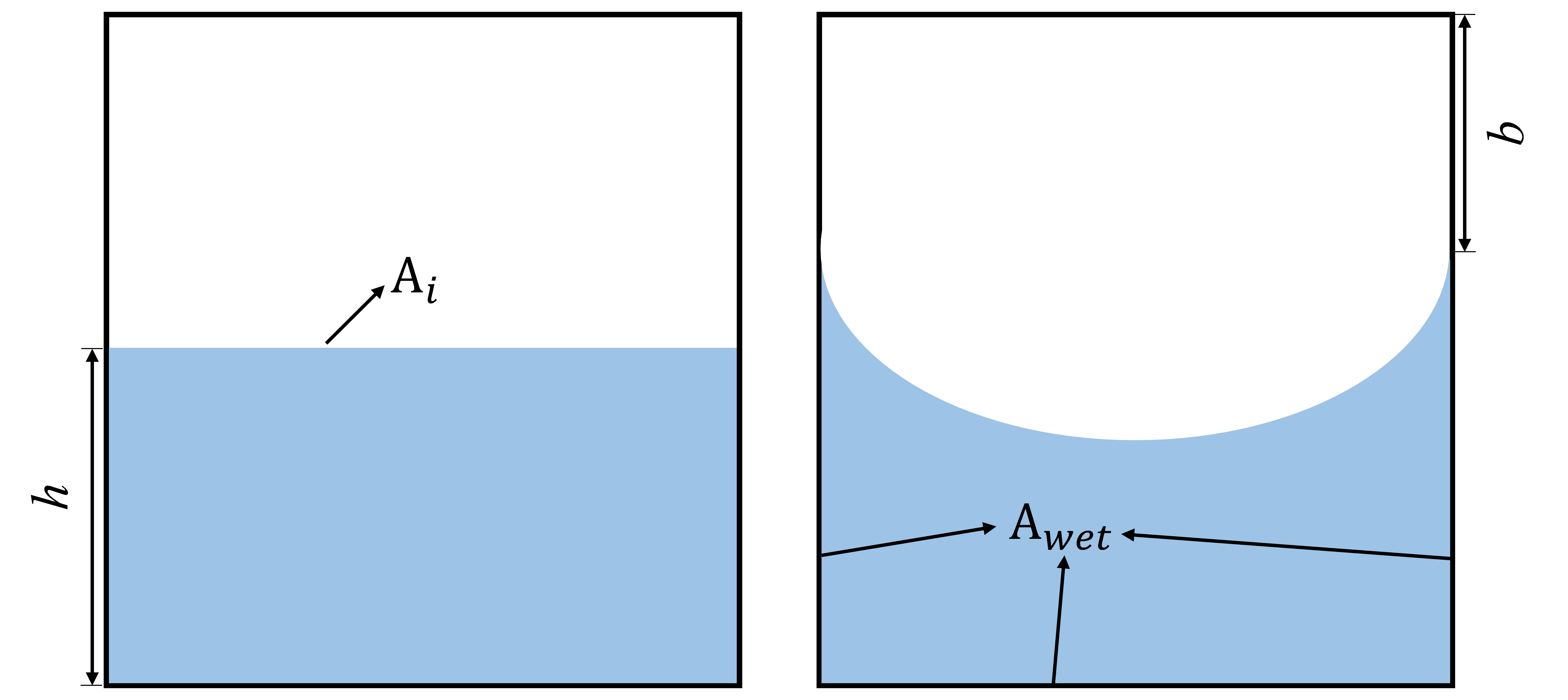}
\caption{Flat interface and spherical interface}
\label{flatvssph}
\end{figure}

Consider a cylindrical tank similar to that used in the simulation as shown in Fig.\ref{flatvssph}, with fill level = 50\%, 
$\theta_c = 0$, $h=1\mathrm{m}$. The liquid volumes in two tanks are equal, however, the 
interface have different shapes. Interface in left is flatten, while in right is a half sphere.
 $b$ is calculated to be $0.2126\mathrm{m}$. So the capillary area in each tank can be calculated as,
\begin{equation}
 A_{c1} = \pi \times \left( \frac{a}{2} \right)^2 - \left[ a\pi h + \pi \times \left( \frac{a}{2} \right)^2 \right] = -a\pi 
\end{equation}

\begin{equation}
 A_{c2} = \frac{1}{2}\times 4\pi \left( \frac{a}{2}\right) ^2-\left[ \pi \times \left( \frac{a}{2} \right)^2 + (a-h-b)a\pi \right] = a\pi \times(h -\frac{3}{4}a-b)
\end{equation}

Clearly we have $A_{c1} > A_{c2}$, and it can be proved that capillary area with spherical interface is lowest under this condition. 

Similar results can be found in microgravity, while another criterion which defined as  $A_c' = \sigma dA_c + Mgdh$ should reach its minimum,
 where $\sigma$ is the surface tension, and $g$ is the local gravity.

\begin{figure}[htb!]
\centering
\includegraphics[width=10cm]{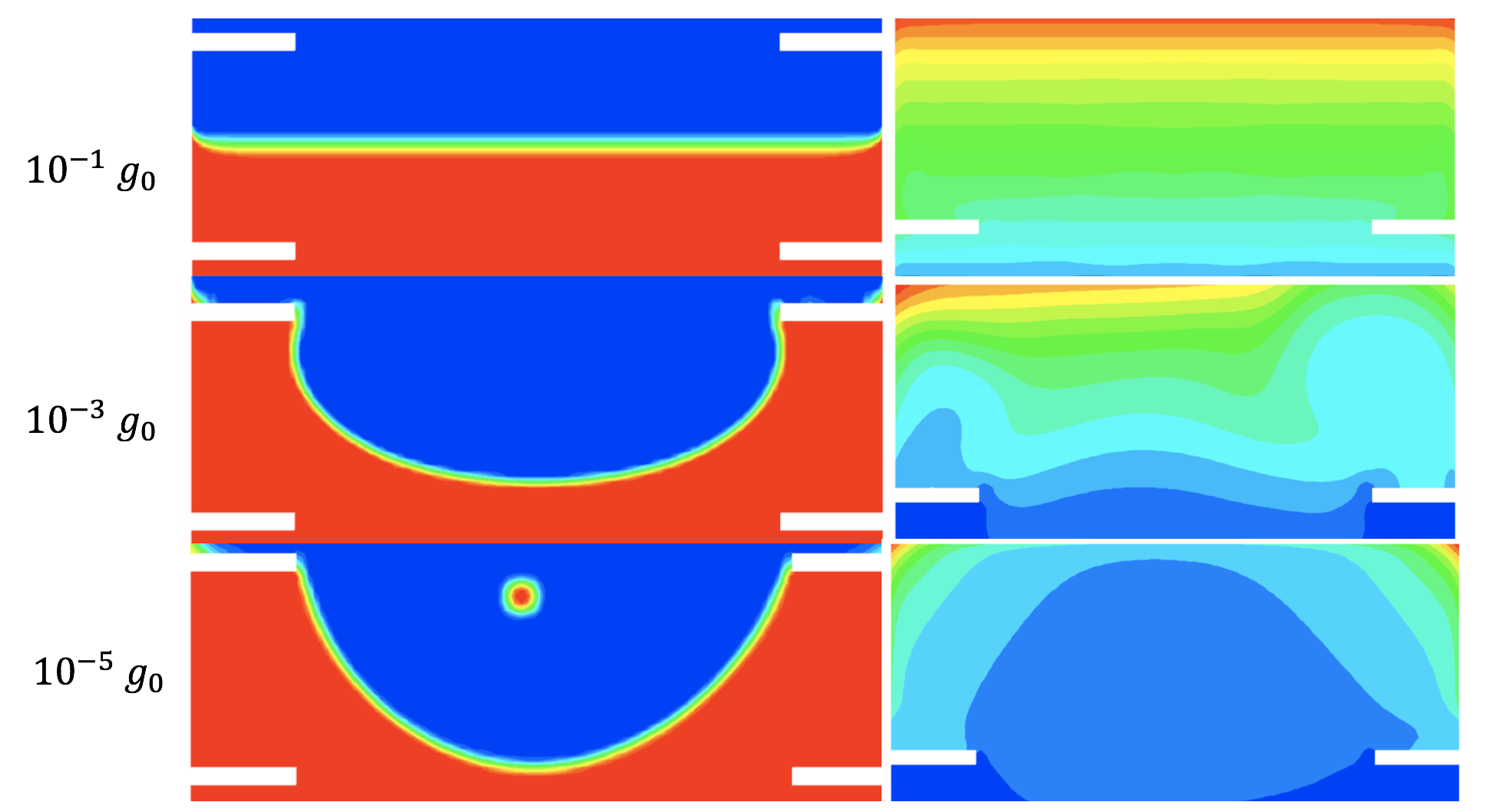}
\caption{Phase and temperature fields under different gravities}
\label{phaseandtemp}
\end{figure}

Figure.\ref{phaseandtemp} shows the phase and temperature distribution under different gravity levels. the curvature of interface and 
temperature layers  decreases as gravity decreases. It can be treated as a lumped element model in high gravity, however, the thermal 
stratification in microgravity will not allow such simplification. $Bi$ number is a dimensionless number which is defined as 
\begin{equation}
Bi = \frac{Lh}{k} 
\end{equation}
where $L$ in this case is average of diameter and height, $h$ is the heat transfer coefficient, $k$ is the thermal conductivity of liquid nitrogen.
$Bi$ in microgravity is larger, which means convection in microgravity
 is more important in space compared to ground.

\begin{figure}[htb!]
\centering
\includegraphics[width=14cm]{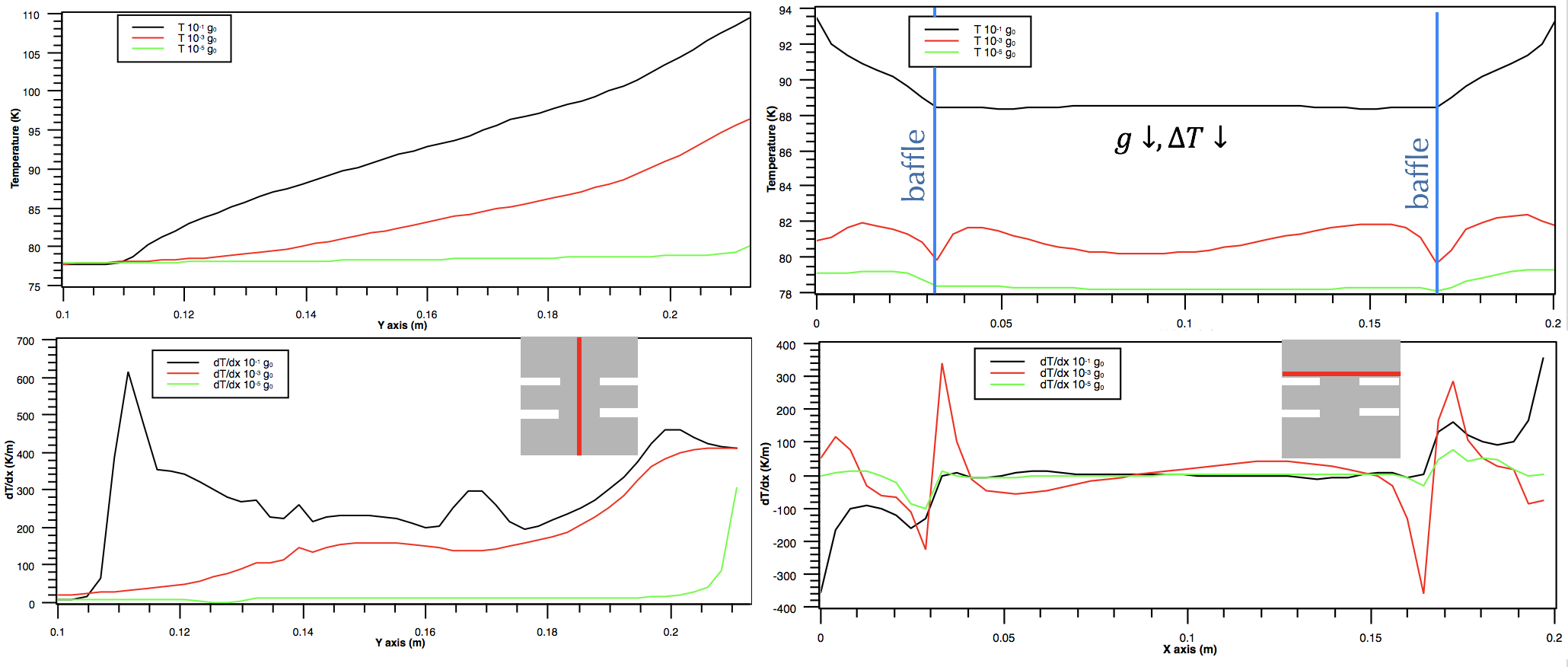}
\caption{Temperature and its gradient along a vertical line(left) and a horizontal line(right) in different gravities}
\label{tempalong}
\end{figure}

Temperature and its gradient in gas phase under different gravities are shown in Fig.\ref{tempalong}. Left side is data along a middle vertical
 line of the tank. Temperature in gas phase is higher in larger gravity. Because in microgravity, the area of liquid wet wall is 
 larger, therefore it takes more time for heat to be transferred to the interface. The temperature derivative in $10^{-5}g_0$ is significantly 
 smaller, and remains a low value until get close to the top tank wall, which is directly heated. 

Right side of Fig.\ref{tempalong} is data along a horizontal line near the upper baffle. The temperature range reduces as the gravity 
decreases. Due to the relatively low temperature rise, the temperature range both in the whole tank and along the baffles are smaller in microgravity. 

\subsection{Fill level}

\begin{figure}[htb!]
\centering
\includegraphics[width=10cm]{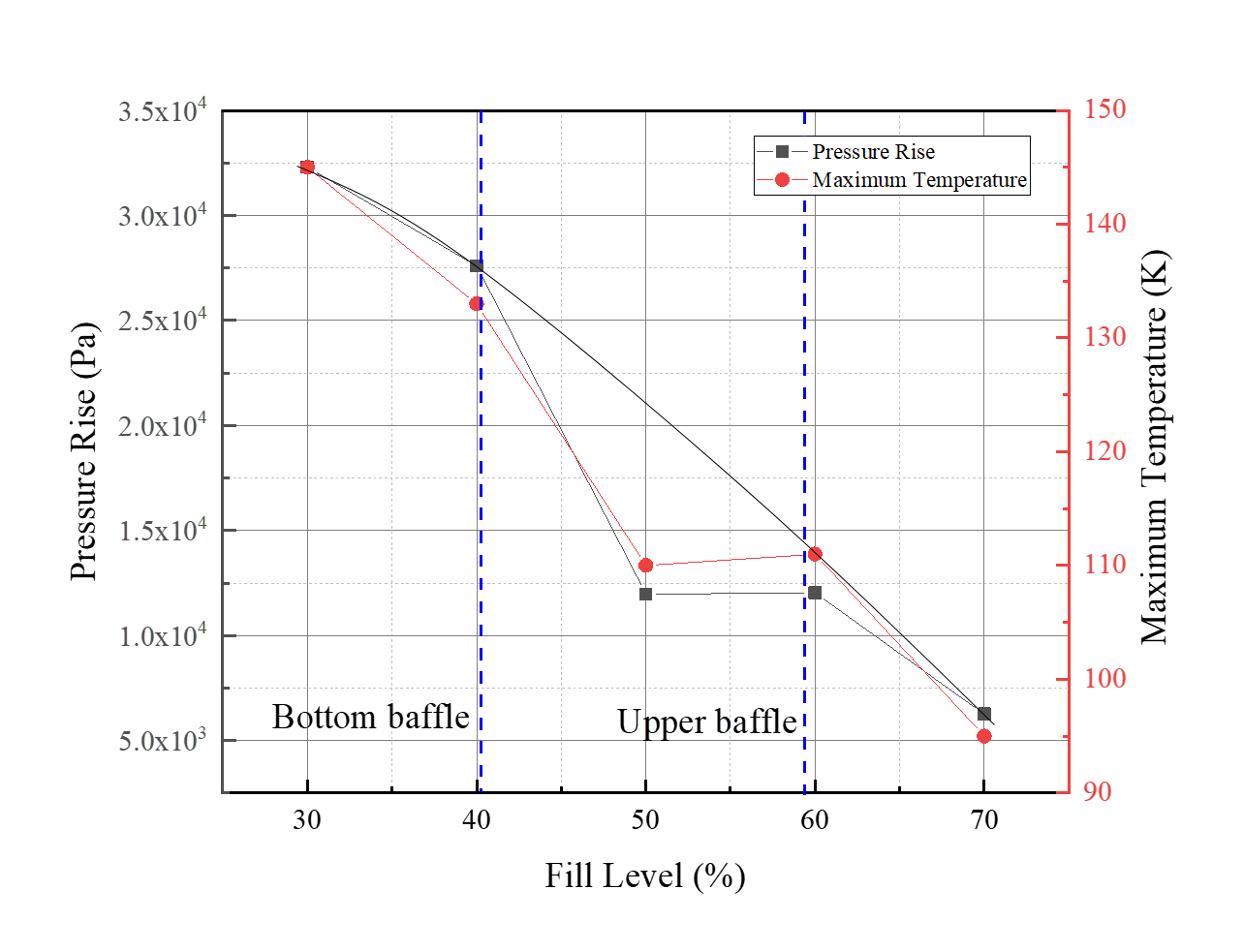}
\caption{Pressure and temperature in different fill levels}
\label{filllevel}
\end{figure}

Pressures and maximum temperatures in $10^{-3} g_0$ are measured after self-pressurization of 1000s. The pressure and temperature change is shown in Fig.\ref{filllevel}.

The pressure rise/temperature versus fill level should has a trend as the dash line in the tank without a baffle. Naturally, the heat leakage helps increase the liquid temperature, promote evaporation and raise pressure in the tank. And the heat flux partitioning model of Kurul and Podowski\cite{kurul1990multidimensional} states that the heat from wall to inner tank consists of three parts, the heat for single phase convective, the heat for quenching and the heat for evaporation. In this case, heat transferred from liquid wet wall first heats the 
nearby liquid nitrogen. Thus the wet wall area and thickness of liquid near wall, which correspond to the ability of blocking heat transfer by liquid, influence pressure and temperature greatly.

\begin{figure}[htb!]
\centering
\includegraphics[width=5cm]{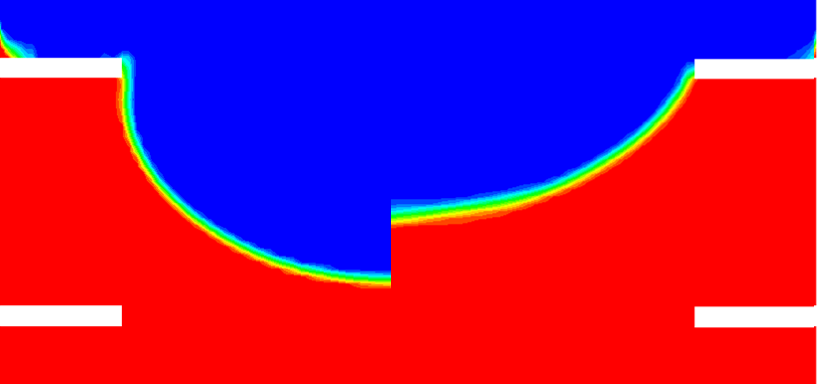}
\caption{Interface of 50\%(left) and 55\%(right) fill levels}
\label{5055}
\end{figure}

Because of the existence of baffles, when fill level increases, liquid will climb up to the edge of baffles by surface tension, thus blocking heat 
transferred to gas phase. Thus the pressure rise will be smaller
if baffles is functioning. So we get a sharp reduction as fill level increases. The transition point is a little higher than 40\%, which can submerge the bottom baffle in normal gravity,
at about 50\%. When fill level is 50\%, the liquid is 
enough to fill the space between two baffles, and liquid filled later will appear at the middle of the tank, where is not effective for blocking heat as wall area. Thus for levels slightly above 50\%, interfaces is similar near wall as shown in the Figure.\ref{5055}, so does the pressure. 
And when fill level is very low, 30\% for example, liquid mainly exists under the bottom baffle and is not enough to cover the wall  
between two baffles. When fill level is above 70\%, the gas ullage exists only in upper section above the upper baffle, thus the effect of baffles is weak. Hence the baffles can mitigate pressure rise for fill levels between two baffles, which is helpful  because the propellant in the tank usually reduces 
from full to empty  during the mission.

\subsection{Distance}
\begin{figure}[htb!]
\centering
\includegraphics[width=9cm]{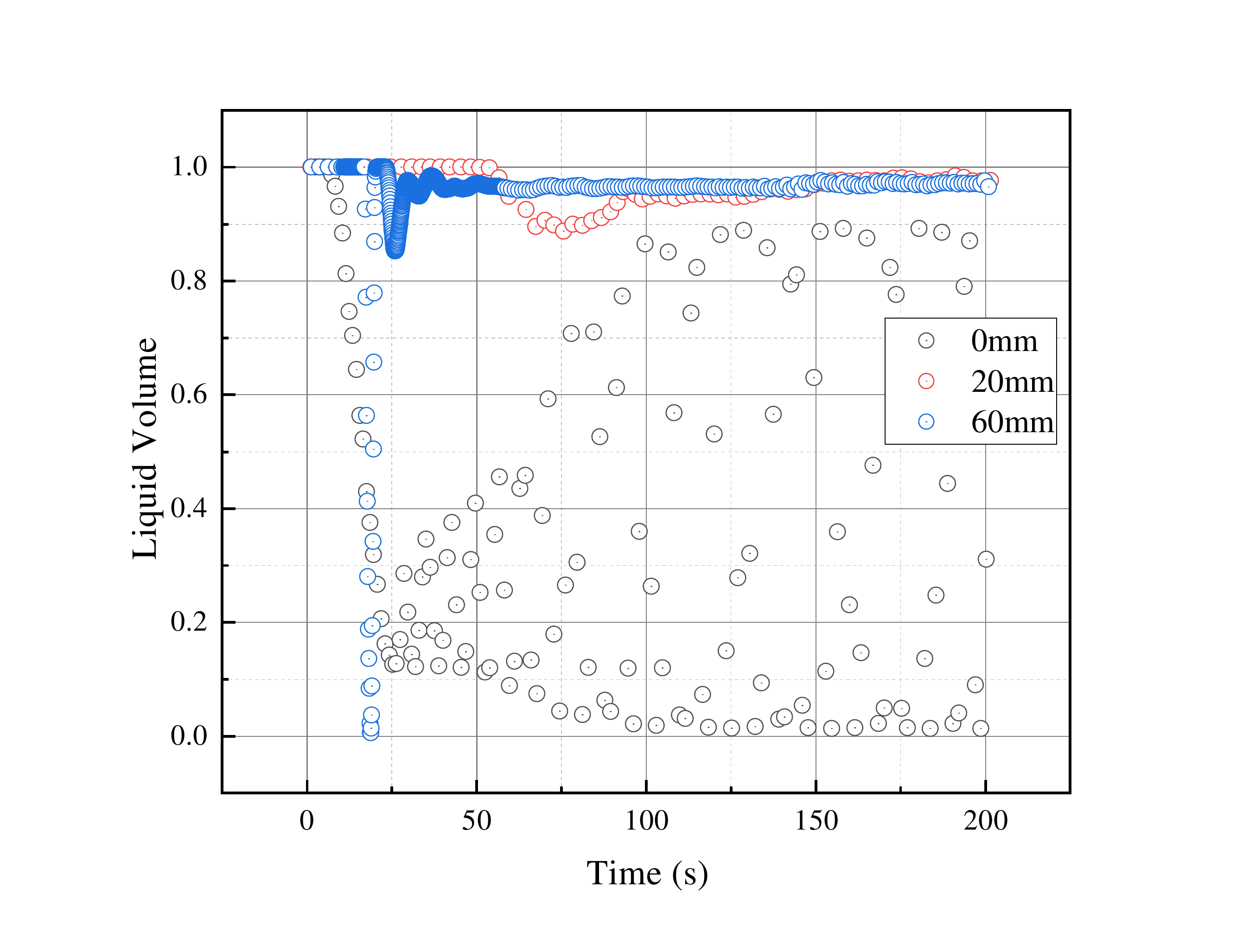}
\caption{Liquid volume near interface in different baffle distances}
\label{fluctuate}
\end{figure}

\begin{figure}[htb!]
\centering
\includegraphics[width=9cm]{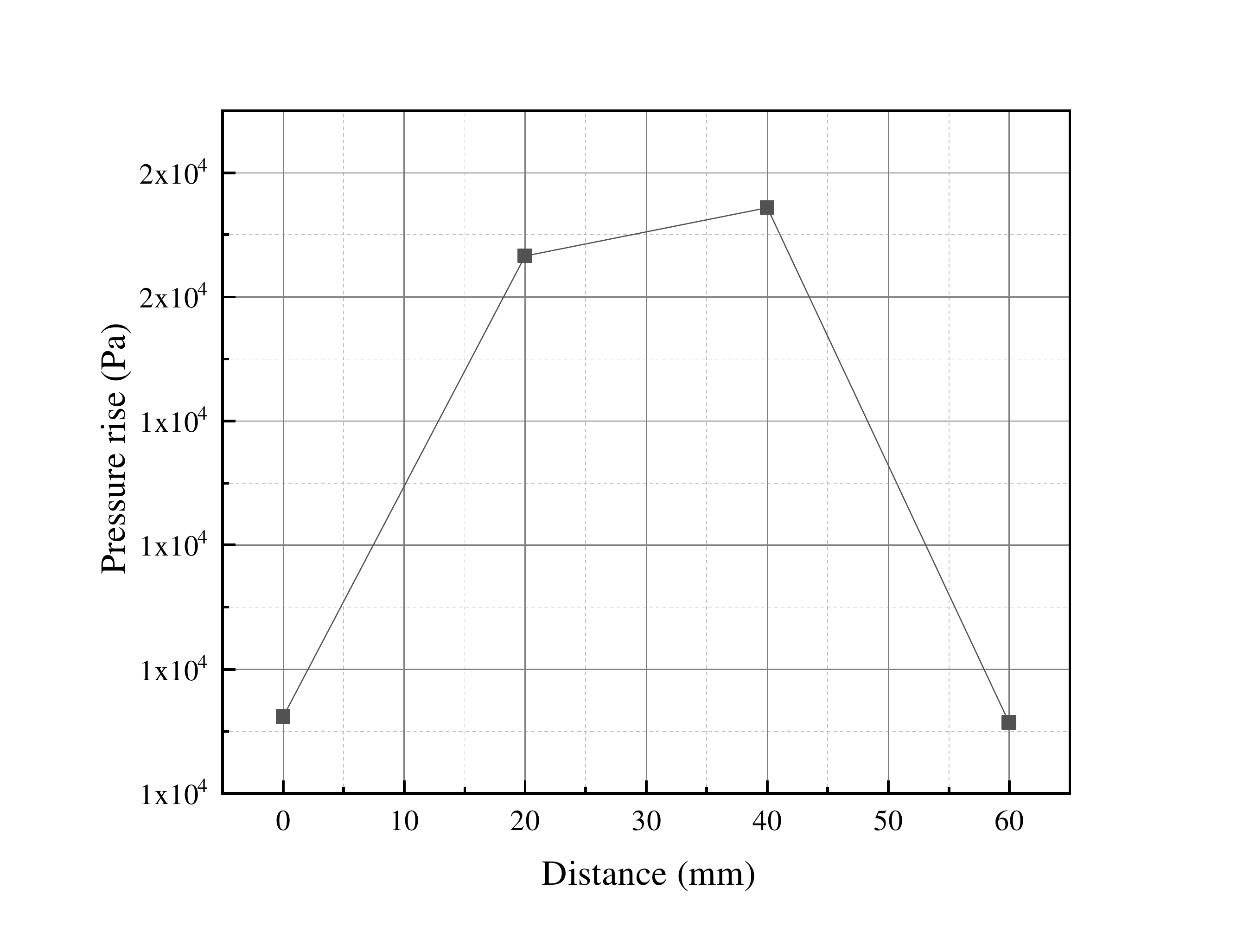}
\caption{Pressure rise at different distance after heated 1000s}
\label{distance}
\end{figure}

Both low and large distances between baffles can mitigate pressure rise, and the curve of pressure rise against distance is shown in Fig.\ref{distance} with fill level 0f 50\% and gravity of $10^{-3} g_0$. 
This is because at distance of 0, representing a single baffle, which is not outstanding in mitigating sloshing, the fluid fluctuates violently, thus the liquid and gas mix a lot. 
As shown in Fig.\ref{fluctuate}, the liquid volume means the liquid fraction in the cell, calculated as,
$\phi = \rho/\rho_l$,
where $\rho$ is the local density and $\rho_l$ is the density of liquid nitrogen. Due to the characteristic of VOF method, liquid volumes near interface have decimal values. The change in liquid volumes
represents movement of interface. After the reorientation process in the first 30s, interface in single baffle tank experiences fast and violent fluctuate. However, as distance increases, the fluctuate is significantly eased. We can see small fluctuates in 
20mm's case, but in 60mm's case, the liquid volume is almost a constant close to 1.
Mixing is an effective method used widely to reduce pressurization in space. By mixing, vapor with higher temperature will condense at splashed liquid. 
The thermal stratification in a thin layer near the liquid-gas interface, which is called thermal boundary layer, will be disturbed. Uniform temperature 
field can be achieved after sufficient mixing. Although temperature of liquid rise a little, the pressure in the tank will decrease. 

\begin{figure}[htb!]
\centering
\subfigure[Phase distribution]{
    \label{pd2060}
    \includegraphics[height=4cm]{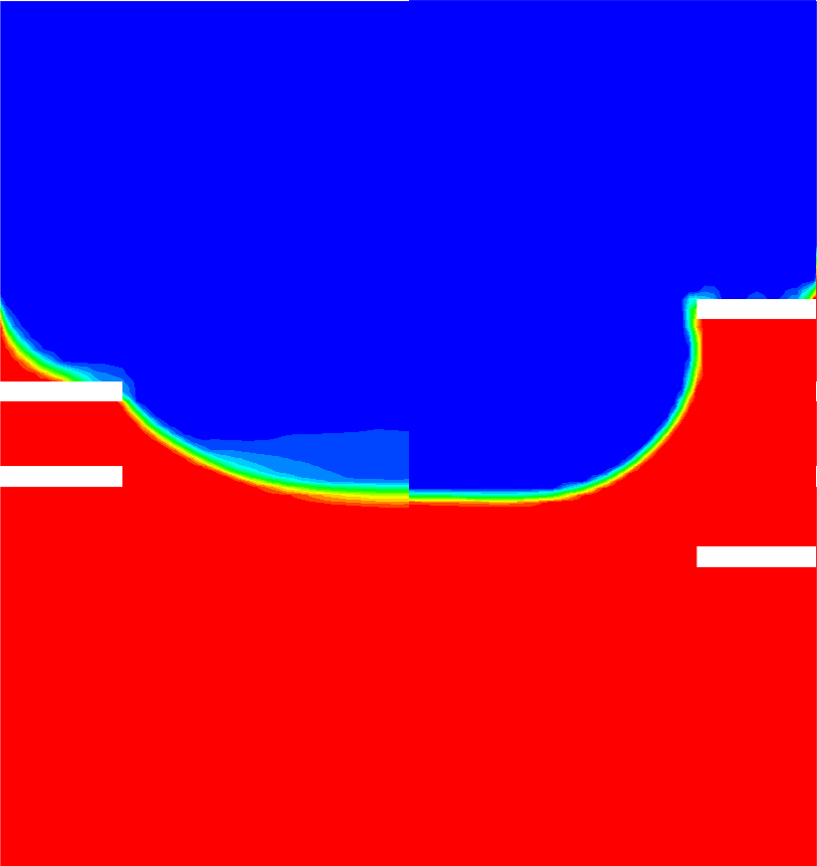}
}
\subfigure[Velocity field]{
    \label{vf2060}
    \includegraphics[height=4cm]{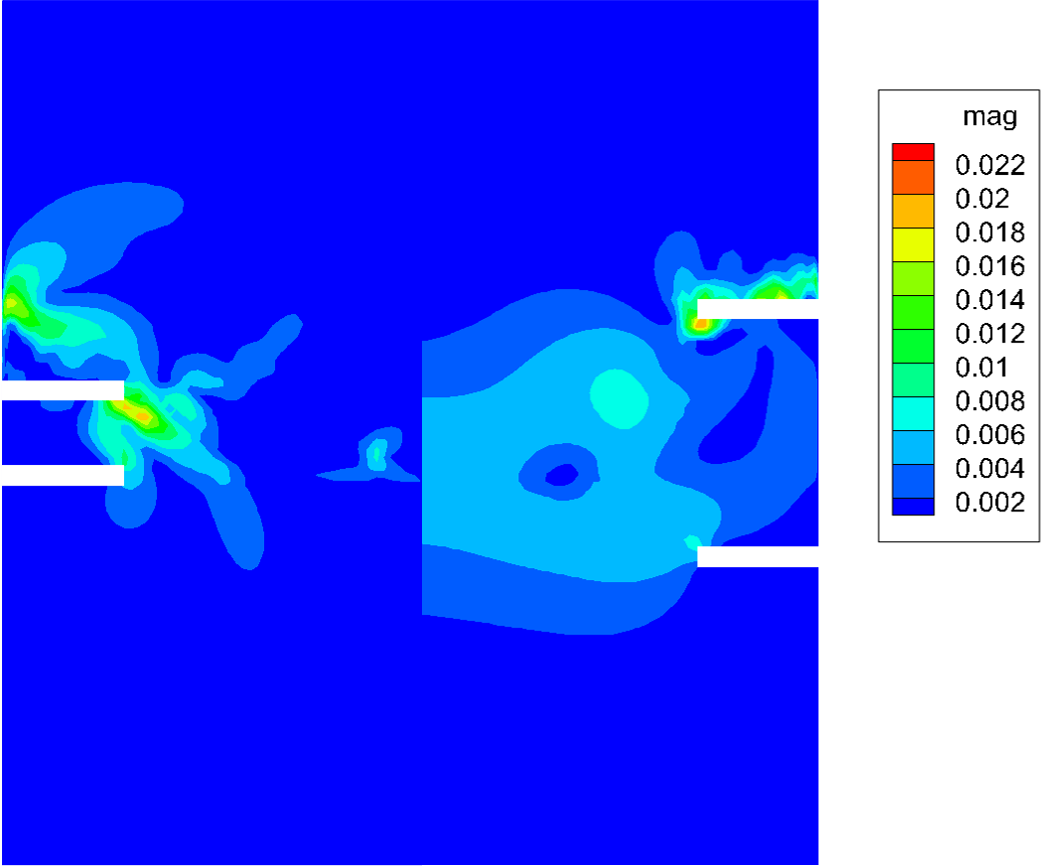}
}
\caption{Phase and velocity field in distances of 20mm and 60mm}
\label{2060}
\end{figure}

As distance increases, the contribution of reducing pressurization changes from mixing to blocking heat transferred into gas as the wet area increases. However,
 in cases that distances are 20mm and 40mm, the mixing phenomenon is weaken by baffles, but the liquid near the wall is too thin to block heat effectively. 
The comparison of phase distribution and velocity field between distances of 20mm and 60mm is shown in Fig.\ref{2060}. So we get the curve as shown in 
Fig.\ref{distance}, in which the final pressure rise after heated 1000s first increases and then decreases after reaching its maximum as the increase of distance.

However, when using single baffle, as the liquid is fluctuating, a periodical force is applied to the tank wall, which is not favorable in practice, as 
some missions, docking for example, are sensitive to the forces in the vehicle, where a quite small force that cannot be correctly predicted will cause 
the failure of the mission. Therefore large distance is preferred when designing in aspect of reducing pressure rise.

\subsection{Orientation}

Simulations are performed for tanks with different baffle orientations. During the simulation, liquid-gas interface climbs up to the edge of baffles at the beginning, then
cover both wall and baffle as much as it can by both surface tension and residual gravity, and then become stable for the rest of the simulation. 

\begin{figure}[htb!]
\centering
\subfigure[Velocity magnitude in the tank]{
    \label{olr}
    \includegraphics[height=4.5cm]{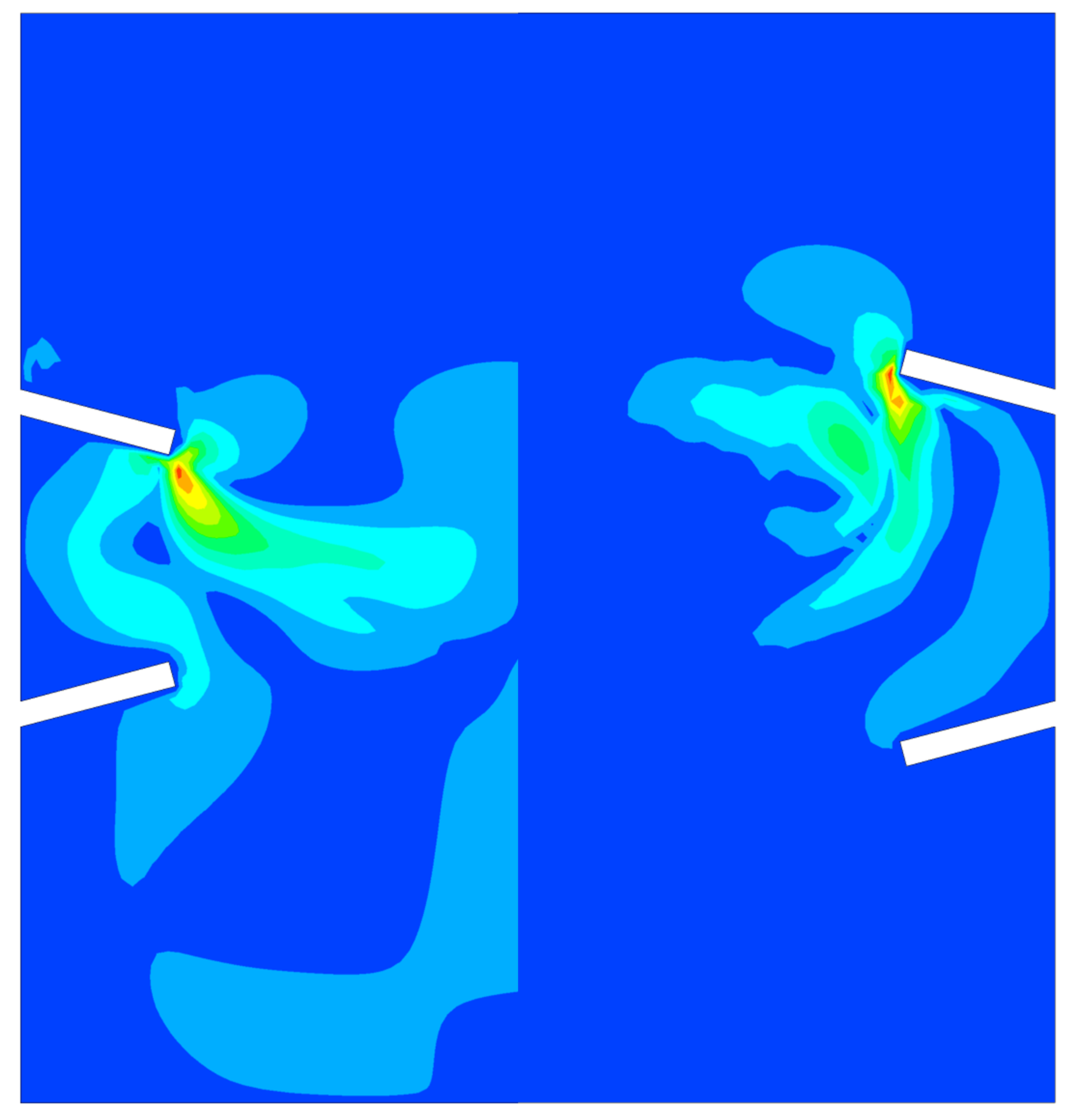}
}
\subfigure[Temperature field of different orientation]{
    \label{ooo}
    \includegraphics[height=4.5cm]{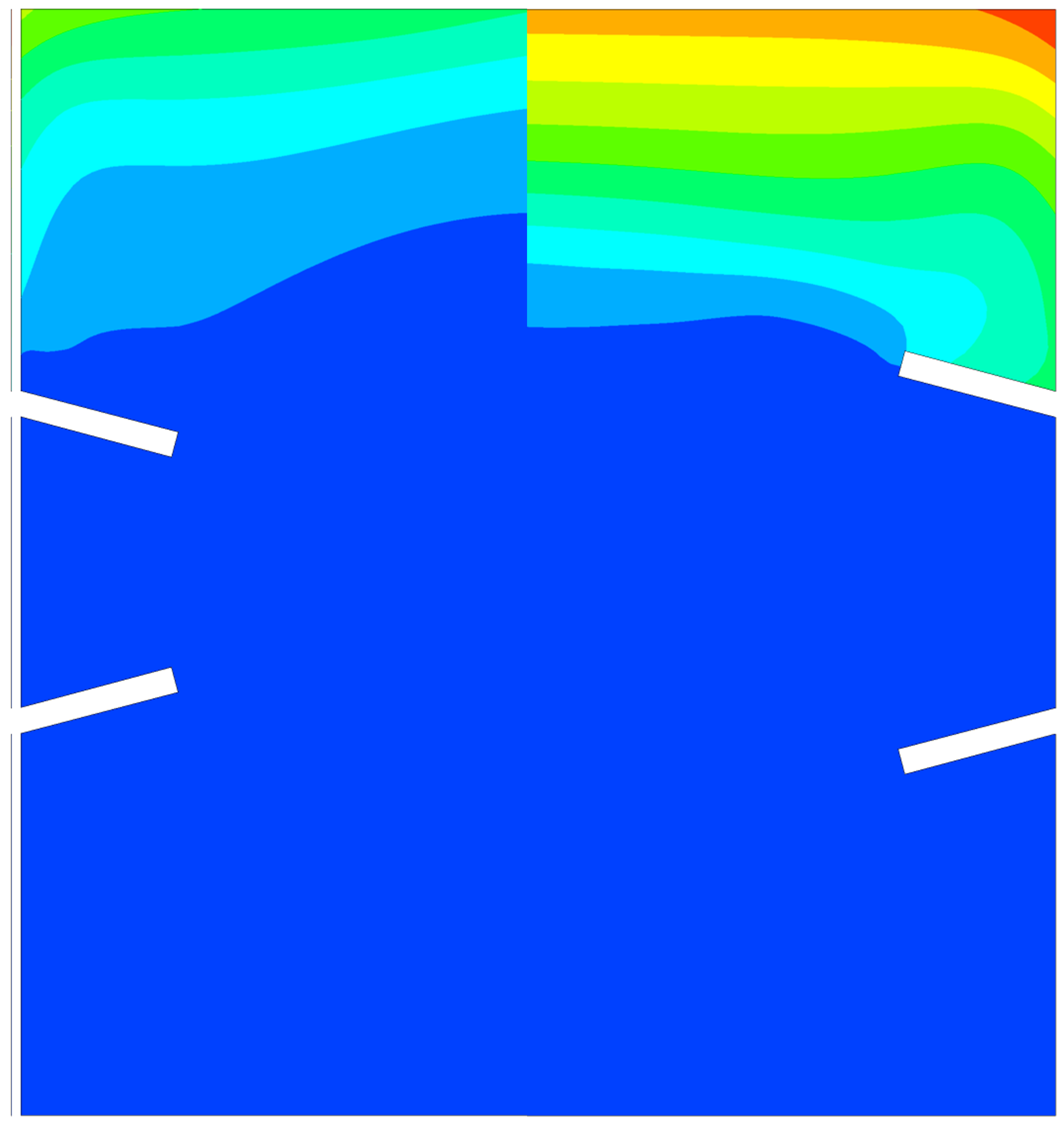}
}
\caption{Velocity and temperature field in tank with different baffle orientation}
\label{olrooo}
\end{figure}

For opposite configuration, $\theta_1$ and $\theta_2$ as shown in Fig.\ref{diagram} is $105^{\circ}$ and $75^{\circ}$ respectively. And for toward configuration, 
$\theta_1$ and $\theta_2$ is $75^{\circ}$ and $105^{\circ}$ respectively. Orientation affects pressure/temperature by affecting the velocity field, which can 
be seen in Fig.\ref{olr}, both figures share the same legend. When the baffles are installed opposite, advection mainly happens between two baffles,
 thus the heat gathers in gas phase and has no path for convection with liquid phase. However, as shown in Fig.\ref{olr}, when the baffles are 
toward, the advection has a wide range, and the average magnitude of velocity is higher. This causes about 10K temperature difference and 6 kPa of pressure 
difference of these two configurations as shown in Table.\ref{ot}. 

Although the maximum temperature in the tank is different in this two configuration, The temperature field in the middle and bottom parts of the tank is similar. 
This is because, despite of the low average velocity in opposite configuration, the convection near baffle is still strong. Thus the temperature in the middle of
 the tank is similar to the other case. However, the convection near the top wall is so weak that cause the difference in maximum temperature difference as 
 discussed above. Difference between toward setup and parallel setup is small, and the velocity field is similar between these two configurations.

During the heating of the tank, it can be found clearly that the baffles can cut the temperature contours, as shown in Fig.\ref{ooo}. There is an apparent 
temperature difference between two side of baffles. This is because of the existence of liquid nitrogen in one side of baffle. In opposite configuration, we can 
see dense contours of temperature end at the end of baffles. 
\begin{table}[htb!]
\centering
\begin{tabular}{cccc}
\hline
 & Toward & Parallel & Opposite \\
 \hline
 $T_{max}(\mathrm{K})$ & 100 & 101 & 114 \\
 $P_{rise}(\mathrm{kPa})$ & 10.74 & 11.60 & 16.54\\
 \hline
 \end{tabular}
 \caption{Temperature and pressure rise in different orientation}
 \label{ot}
 \end{table}

\subsection{Gap}

\begin{figure}[htb!]
\centering
\subfigure[gap=0mm]{
    \label{g1}
    \includegraphics[height=4cm]{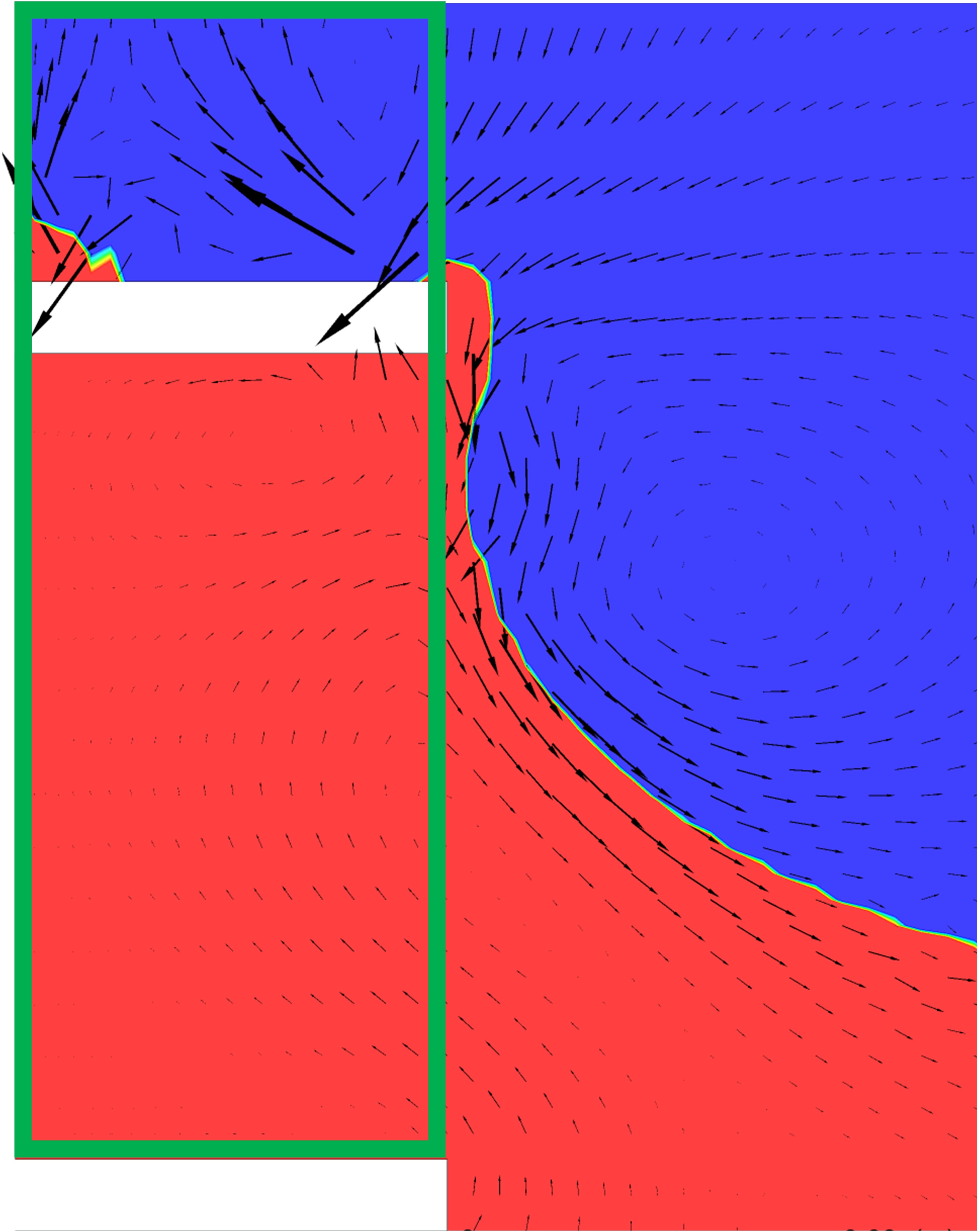}
}
\subfigure[gap=5mm]{
    \label{g2}
    \includegraphics[height=4cm]{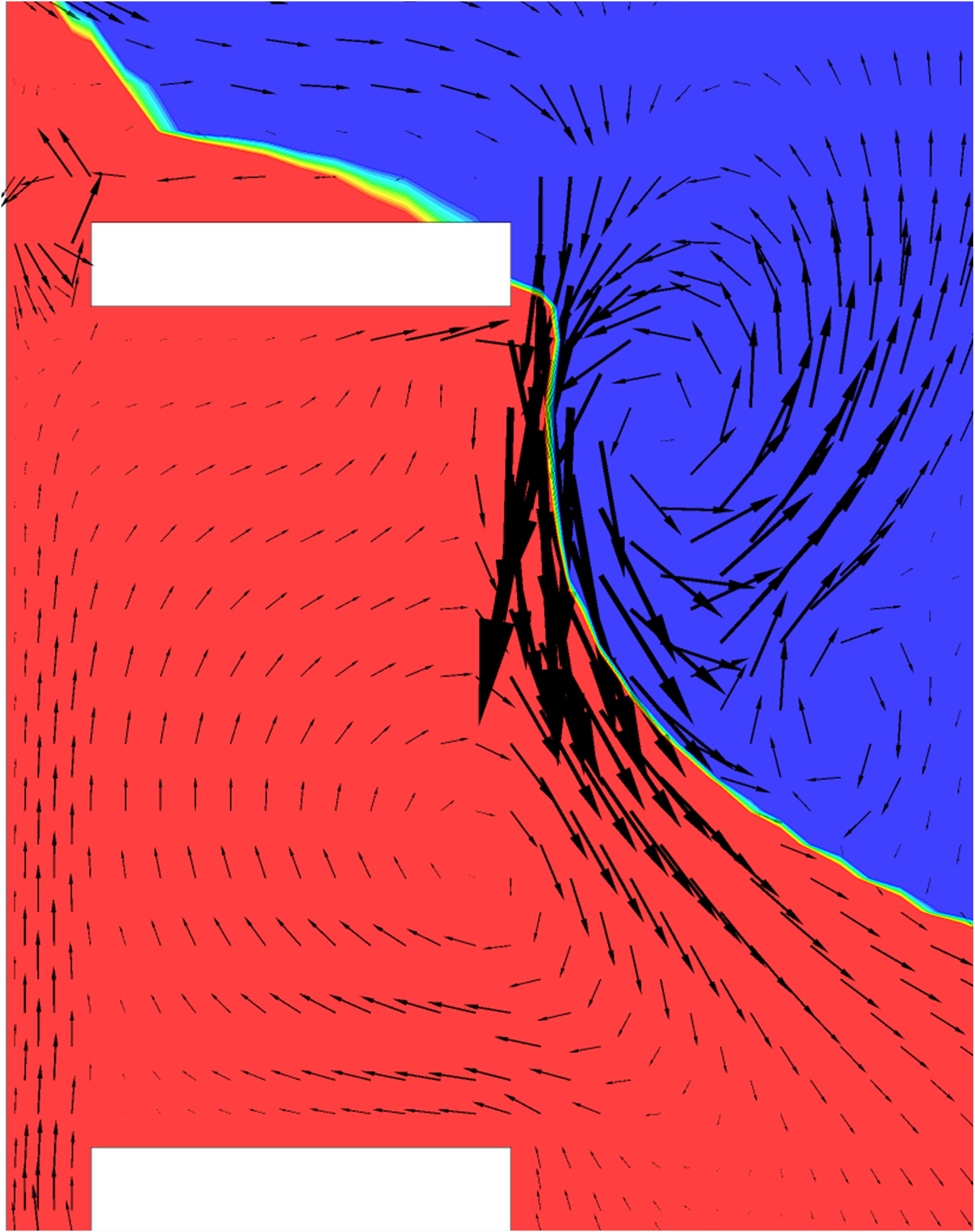}
}
\subfigure[gap=10mm]{
    \label{g3}
    \includegraphics[height=4cm]{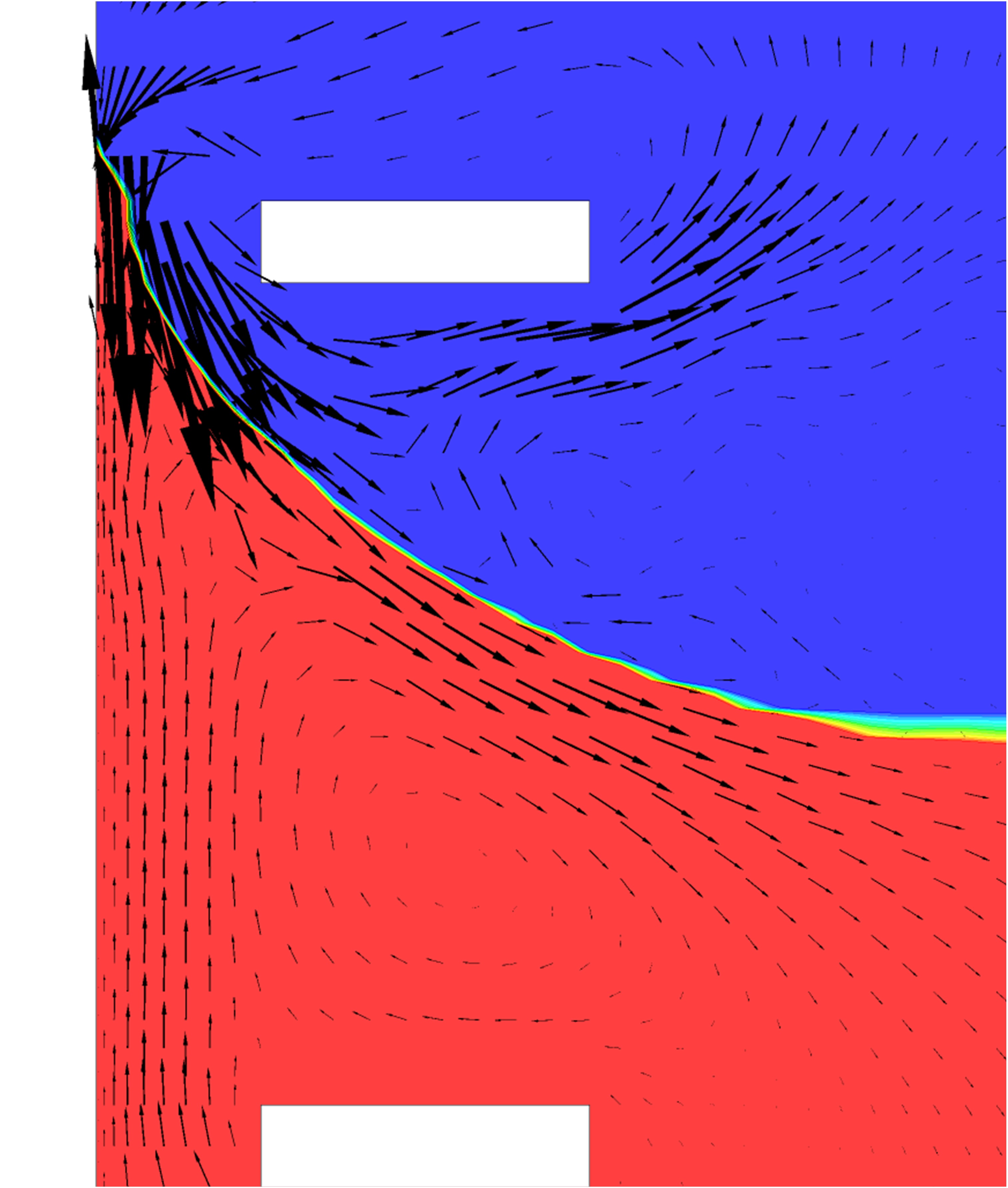}
}
\caption{Phase and velocity vector of different gap between baffles and wall}
\label{ga}
\end{figure}

Fig.\ref{ga} shows the phase distribution and velocity vector in the tank with different gaps between baffles and wall, where liquid phase is blue 
and gas phase is yellow. The gap results from the holes on the baffle, which kind of baffle is adopted in many researches discussing reducing 
the weight of baffles\cite{yoon2015effect}. In this research, the total length of baffle is kept constant. The heat flux is relatively small compared to 
occasions where bubbles generates, and the phase transfer is mainly happens near the interface. Vortex generates under the effect of evaporation, 
which is similar to normal gravity. The location of maximum velocity is sightly under the upper baffles.

\begin{figure}[htb!]
\centering
\includegraphics[width=9cm]{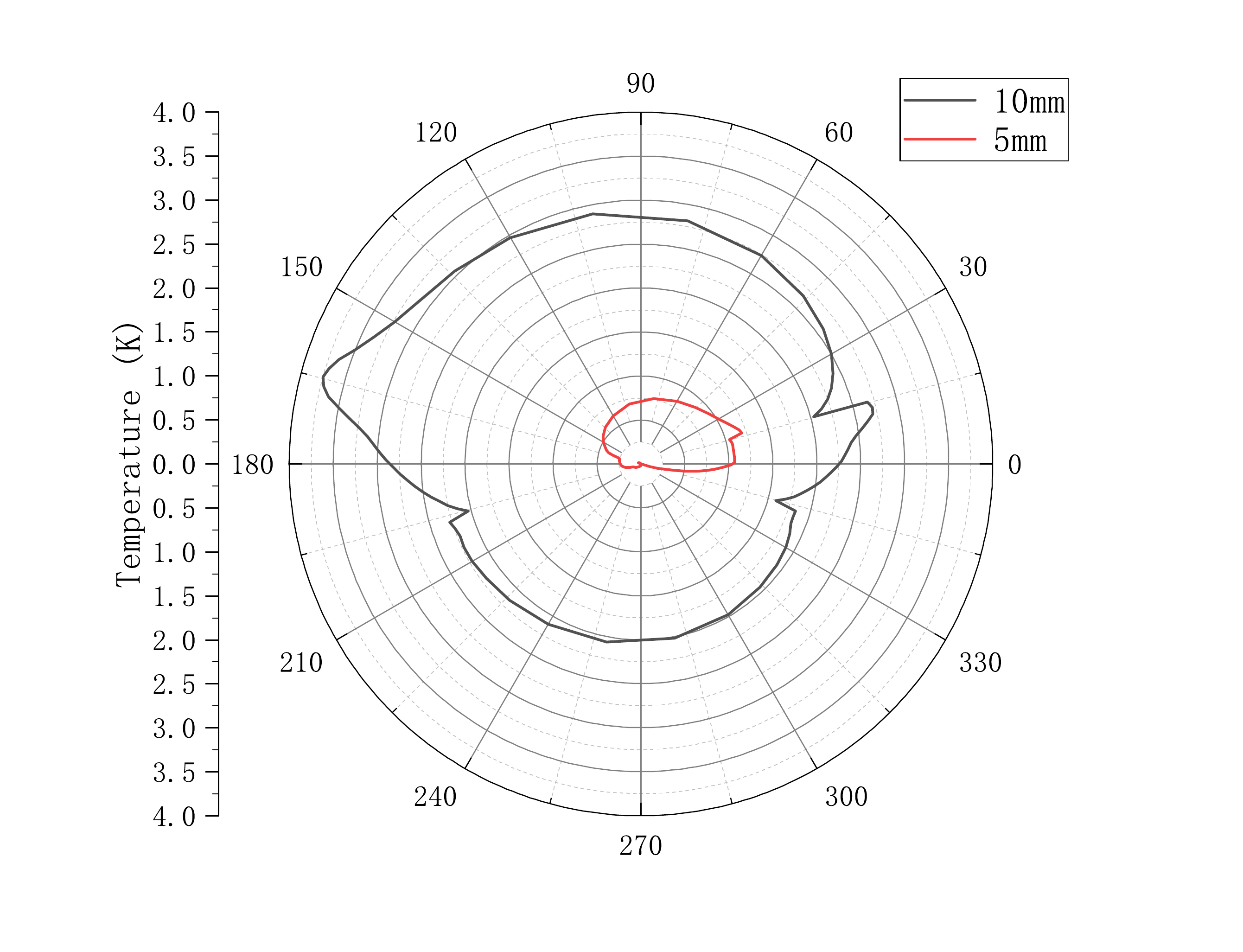}
\caption{Temperature distribution around the baffle}
\label{hole-tem}
\end{figure}

 Baffle alert the direction of saturated airflow around it, resulting high temperature region as mentioned in Ref.\cite{grayson2006cryogenic}. 
 A comparison of temperature distribution around the upper left baffle is shown in Fig.\ref{hole-tem}, the geometric center of the baffle is set 
 to be the origin point in the polar. The temperature has been subtracted 77.4K in plot. The maximum temperature 
 occurs at the upper left corner of the baffle with 10mm's gap, however, the maximum temperature with 5mm's gap occurs at upper right corner. This
  is because with a small gap, 5mm for example, the liquid will climb above the baffle by surface tension, the left side of the baffle with 5mm's
   gap is full of low-temperature liquid nitrogen. However, in 10mm's gap case, the left side of the baffle is the high-temperature evaporated nitrogen
    gas flow. This also causes 2.0K of difference in average temperature along the baffle of these two cases, as shown in the figure. 
    The pressure rise in case of 5mm gap is minimum. The behavior of interface in case of 10mm gap is similar to situations without a baffle, and the 
pressure is largest in this case.

\section{CONCLUSION}

\begin{figure}[htb!]
\centering
\includegraphics[width=9cm]{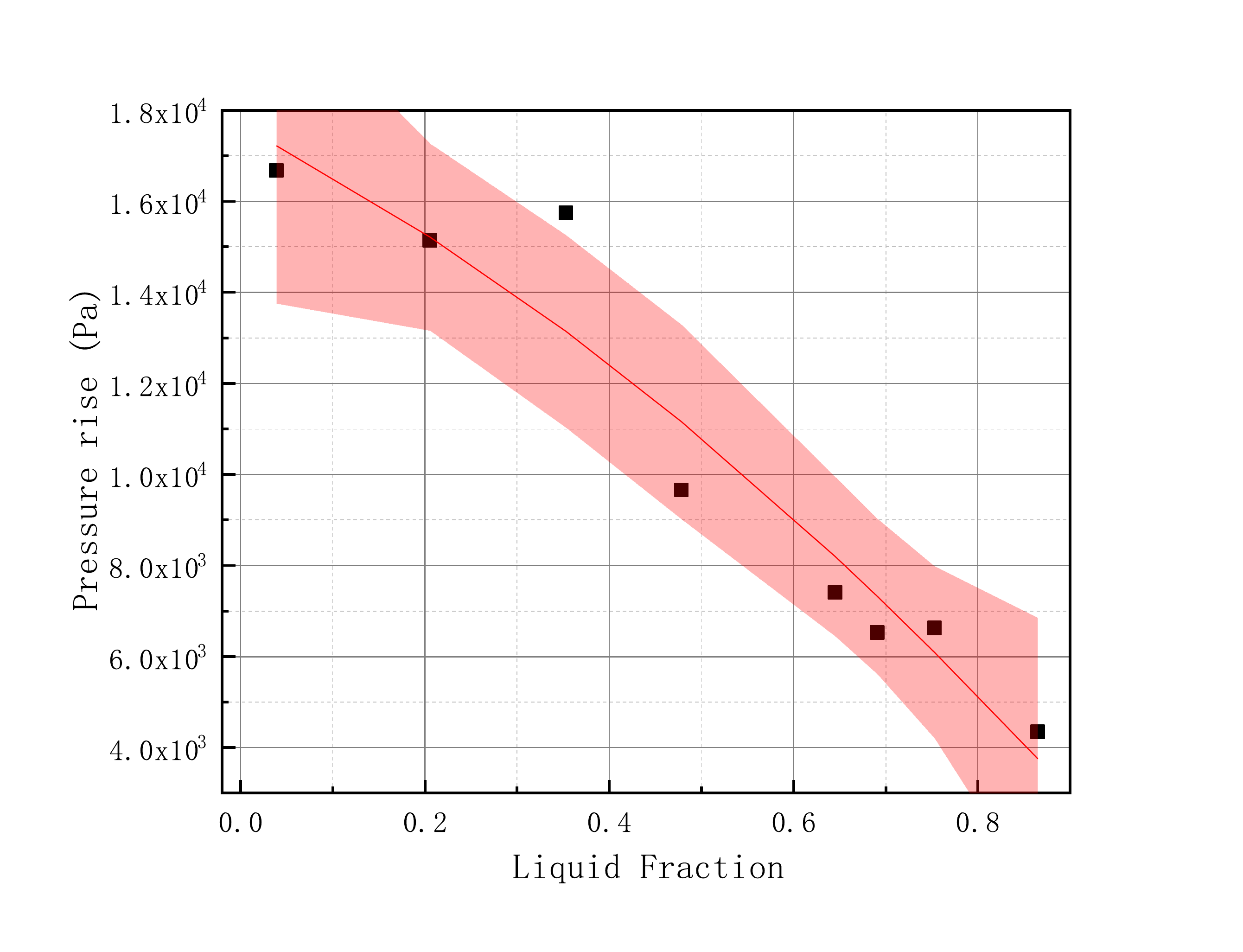}
\caption{Relationship between pressure rise and liquid fraction}
\label{rplf}
\end{figure}

From the discussion above we can see that the liquid near the wall plays an important role in affecting the 
pressure rise in the tank. In fill level section, the fill level affects the wet wall fraction directly; In gap section, the
existence of gap influences the wet wall area because surface tension cannot be neglected in space, and liquid can climb up 
to the baffle if the gap is small. Efforts to quantitatively describe the effect of these two factors are made, and a specific region,
which is shown in \ref{g1} as the green rectangle, is chosen to describe the pressure rise between fill level of 30\% and 70\%. The top of 
rectangle is 3cm higher than the upper baffle, the bottom of the rectangle coincide with the bottom baffle, and the width of the rectangle
is same with the width of baffles. The fraction 
of liquid area in this rectangle is plotted in Fig.\ref{rplf} with the pressure rise in the tank after heated 300s. The data includes cases for different
fill levels, gaps and distances, which are all for gravity of $10^{-3} g_0$.

The pressure rise decrease with the increase of the liquid fraction clearly. A polynomial fitting is shown in Fig.\ref{rplf}, and the function writes as,

\begin{equation}
P = 17640 -10453 x - 6480 x^2 
\end{equation}

where $P$ is the pressure rise in the tank, and $x$ is the liquid fraction. When $x=0$, the liquid is all beneath the bottom baffle, and baffles have no 
influence on the behavior of liquid nitrogen. When $x=1$, two baffles are submerged in liquid nitrogen totally, and because the tank is steady, baffles 
can do little with liquid nitrogen. Almost every point stand in 95\% to 105\% of the fitting curve, except the data from fill level of 45\%. This probably
because 45\% is slightly smaller than the transition point discussed in Section 3.3, and in this case, the wet wall area is large but liquid
nitrogen near the wall is thin. This implies that the thickness of liquid may have more important role in reducing pressurization than wet area.

 In this paper, we focused on the influences of parameters of baffles to thermal-stratification and pressurization in the tank in various 
 conditions. We discussed the effect of gravity levels and fill levels, which corresponding to different tasks' environments and periods. 
 The results shows that as the gravity reduces, the thermodynamic behavior of liquid will differ from it in normal gravity. The pressure change
  will be small, and the derivative of pressure is small too. And as the reduction of fuel from full to empty, the pressure will rise more rapidly.
   The baffles can mitigate pressure rise when the fill level is between the two baffles. 

  Then we focused on the distances, orientations, and gaps of the baffles, which are all problems during design process. Clearly, These parameters
   have a significant effect on thermodynamical distribution in the tank under microgravity. The effect comes from differences in velocity field, phase distribution and etc.

  Baffles play different roles as working conditions or its shapes change, and by optimizing the parameters of baffles, up to 54\% of reduction in
   pressure rise can be achieved (by changing the orientation from opposite to toward). Higher performance can be easily achieved by coupling two
    or more factors discussed before. Efforts for designing tanks can be saved by realizing the important role of baffles in reducing pressurization under microgravity, 
    especially for long duration space missions.

\bibliographystyle{unsrt}
\bibliography{arxiv}

\begin{thebibliography}{10}

\bibitem{panzarella2003validity}
Charles~H Panzarella and Mohammad Kassemi.
\newblock On the validity of purely thermodynamic descriptions of two-phase
  cryogenic fluid storage.
\newblock {\em Journal of Fluid Mechanics}, 484:41--68, 2003.

\bibitem{sumner1966experimental}
Irving~E Sumner, Raymond~F Lacovic, and Andrew~J Stofan.
\newblock {\em Experimental Investigation of Liquid Sloshing in a Scale-Model
  Centaur Liquid-Hydrogen Tank}.
\newblock National Aeronautics and Space Administration, 1966.

\bibitem{chintalapati2008parametric}
Sunil Chintalapati and Daniel Kirk.
\newblock Parametric study of a propellant tank slosh baffle.
\newblock In {\em 44th AIAA/ASME/SAE/ASEE Joint Propulsion Conference \&
  Exhibit}, page 4750, 2008.

\bibitem{hasheminejad2011effect}
Seyyed~M Hasheminejad and MM~Mohammadi.
\newblock Effect of anti-slosh baffles on free liquid oscillations in partially
  filled horizontal circular tanks.
\newblock {\em Ocean Engineering}, 38(1):49--62, 2011.

\bibitem{hasheminejad2012sloshing}
Seyyed~M Hasheminejad and Mostafa Aghabeigi.
\newblock Sloshing characteristics in half-full horizontal elliptical tanks
  with vertical baffles.
\newblock {\em Applied Mathematical Modelling}, 36(1):57--71, 2012.

\bibitem{yoon2015effect}
Sung-Ho Yoon and Kee-Jin Park.
\newblock Effect of baffles on sloshing mitigation in liquid storage tanks.
\newblock 2015.

\bibitem{dodge2000new}
Franklin~T Dodge et~al.
\newblock {\em The new" dynamic behavior of liquids in moving containers"}.
\newblock Southwest Research Inst. San Antonio, TX, 2000.

\bibitem{kannapel1987liquid}
M~KANNAPEL, A~PRZEKWAS, A~SINGHAL, and N~COSTES.
\newblock Liquid oxygen sloshing in space shuttle external tank.
\newblock In {\em 23rd Joint Propulsion Conference}, page 2019, 1987.

\bibitem{behruzi2014cryogenic}
Philipp Behruzi, Martin Konopka, Francesco de~Rose, Guido Schwartz, et~al.
\newblock Cryogenic slosh modeling in lng ship tanks and spacecrafts.
\newblock In {\em The Twenty-fourth International Ocean and Polar Engineering
  Conference}. International Society of Offshore and Polar Engineers, 2014.

\bibitem{ma2017investigation}
Yuan Ma, Yanzhong Li, Kang Zhu, Ying Wang, Lei Wang, and Hongbo Tan.
\newblock Investigation on no-vent filling process of liquid hydrogen tank
  under microgravity condition.
\newblock {\em International Journal of Hydrogen Energy}, 42(12):8264--8277,
  2017.

\bibitem{adam2014design}
Patrick Adam and Jacob Leachman.
\newblock Design of a reconfigurable liquid hydrogen fuel tank for use in the
  genii unmanned aerial vehicle.
\newblock In {\em AIP Conference Proceedings}, volume 1573, pages 1299--1304.
  AIP, 2014.

\bibitem{grayson2006cryogenic}
Gary Grayson, Alfredo Lopez, Frank Chandler, Leon Hastings, and Stephen Tucker.
\newblock Cryogenic tank modeling for the saturn as-203 experiment.
\newblock In {\em 42nd AIAA/ASME/SAE/ASEE Joint Propulsion Conference \&
  Exhibit}, page 5258, 2006.

\bibitem{seo2010analysis}
Mansu Seo and Sangkwon Jeong.
\newblock Analysis of self-pressurization phenomenon of cryogenic fluid storage
  tank with thermal diffusion model.
\newblock {\em Cryogenics}, 50(9):549--555, 2010.

\bibitem{kartuzova2011modeling}
Olga Kartuzova and Mohammad Kassemi.
\newblock Modeling interfacial turbulent heat transfer during ventless
  pressurization of a large scale cryogenic storage tank in microgravity.
\newblock In {\em 47th AIAA/ASME/SAE/ASEE Joint Propulsion Conference \&
  Exhibit}, page 6037, 2011.

\bibitem{hirt1981volume}
Cyril~W Hirt and Billy~D Nichols.
\newblock Volume of fluid (vof) method for the dynamics of free boundaries.
\newblock {\em Journal of computational physics}, 39(1):201--225, 1981.

\bibitem{kurul1990multidimensional}
N~Kurul and Mm~Z Podowski.
\newblock Multidimensional effects in forced convection subcooled boiling.
\newblock In {\em Proceedings of the Ninth International Heat Transfer
  Conference}, volume~2, pages 19--24. Hemisphere Publishing New York, 1990.

\end{thebibliography}
\end{document}